\documentclass[a4paper,12pt]{article}
\usepackage{epsfig}
\newcommand{\com}[1]{}
\renewcommand{\vec}[1]{\relax\ifmmode\mathchoice
{\mbox{\boldmath$\relax\displaystyle#1$}}
{\mbox{\boldmath$\relax\textstyle#1$}}
{\mbox{\boldmath$\relax\scriptstyle#1$}}
{\mbox{\boldmath$\relax\scriptscriptstyle#1$}}\else
\hbox{\boldmath$\relax\textstyle#1$}\fi}
\begin{document}
\pagestyle{myheadings}
\markboth{K\"olbl/Helbing: Energy and Scaling Laws in 
Human Travel Behaviour}
{K\"olbl/Helbing: Energy and Scaling Laws in 
Human Travel Behaviour}

\title{\Large\bf Energy and Scaling Laws in 
Human Travel Behaviour}
\author{\large Robert K\"olbl$^{1,2}$ and Dirk Helbing$^{3,4}$\\[4mm]
\normalsize $^1$ Institute for Transport Planning and Traffic Engineering,\\
\normalsize University of
Technology Vienna, Gusshausstrasse 30, 1040 Vienna, Austria\\
\normalsize $^2$ %Highfield, 
University of Southampton, Southampton, SO17 1BJ, UK\\
\normalsize $^3$ Institute for
Economics and Traffic,\\ 
\normalsize Dresden University of Technology, 
Andreas-Schubert-Str. 23, 01062 Dresden, Germany\\
\normalsize $^4$ Collegium Budapest~-- Institute for Advanced Study,
%Szenth\'{a}roms\'{a}g u. 2,
H-1014 Budapest, Hungary}
\maketitle
\vfill
{\bf We show that energy concepts can contribute to the understanding
of human travel behaviour. First, the average travel times for different modes of
transportation are inversely proportional to the energy consumption rates measured for
the respective physical activities. Second,
when daily travel-time distributions by different modes of transport
such as walking, cycling, bus or car travel are appropriately scaled, they
turn out to have a universal functional relationship. This corresponds to a canonical-like 
energy distribution with exceptions for short trips, which can be theoretically
explained. Altogether, this points to a law of constant average energy consumption 
by the physical activity of daily travelling. Applying these natural laws 
could help to improve long-term urban and transport planning.
} 
%\pacs{PACS: 04.20.Fy,%Canonical formalism, Lagrangians, and variational principles 
%%05.20.Gg,%Classical ensemble theory
%05.70.-a,%Thermodynamics
%%87.19.St,%Movement and locomotion 
%89.40.+k,%Transportation 
%89.50.+r%Urban planning and development 
%}
\vfill
\clearpage
\section{Introduction}

During the last decade, physicists have discovered the significance of many methods
from statistical physics and nonlinear dynamics for the description of traffic flows
\cite{general,Review,TGF01}, including master and Fokker-Planck equations \cite{master}, 
molecular dynamics and cellular automata \cite{CA}, 
gas kinetics and fluid dynamics \cite{gasfluid}, 
instability and phase diagrams \cite{phase},
complex pattern formation \cite{pattern},
Korteweg-de-Vries and Ginzburg-Landau equations \cite{Ginzburg}, 
and invariants or universal constants of traffic flow \cite{const}.
The observed instability mechanisms, jamming, segregation,
breakdown and clustering phenomena are now viewed as a paradigm for similar phenomena
in granular and colloidal physics \cite{TGF01,TGF99,Colloid}, in biology (social insects),
medicine (evacuation, coagulation), logistics (instability of supply chains), 
and economics (business cycles, optimal production) \cite{Review,TGF01,TGF99,Econo}. The success
of traffic theory is partly based on the fact that it has delivered the first quantitative theory
of a system involving human behaviour. 
\par
Scientists have also tried to apply physical laws to other areas of transportation. For example,
they have transfered the gravity law to fit migration rates and origin-destination matrices 
of streams of persons or goods \cite{gravity}. Moreover, in order to derive more appropriate 
mathematical relations for these, they have applied entropy principles \cite{sechs,entropy}. In fact,
some of the most common decision models such as the so-called multinomial 
logit model \cite{MNL} can be derived from entropy principles \cite{sociodyn}. Presently, 
the discussion on regularities in travel times is dominated by the idea of a constant travel time budget \cite{fuenf},
which is not compatible with entropy-based decision laws. It states that, 
on average, humans use to travel about 75~minutes per day since many
decades (or even centuries). This conjecture is the basis of most studies on induced traffic.
\par
We will show that the concept of a constant travel time budget is only consistent with empirical data,
if different modes of transport are not distinguised. It will turn out to be a
special case of a more general law of a constant travel energy budget, which is proposed in
this paper. This energy-based travel time concept is also consistent with entropy-based
decision theories. In our study, we will first investigate the average travel times over many years
separately for different modes of transport (see Sec.~\ref{Sec1}).  We will find large differences between
the mode-specific travel times, but (considering the small sampling size)
they are relatively stable over 25 to 30 years, i.e. typical
planning periods. Interestingly, the proportions of the mode-specific average travel times
are inversely proportional to the typical energy consumption rates by the body during 
the related travel activities. This is a first support of the idea that energy concepts
may help to understand human travel behaviour. In Section \ref{Sec2}, we will then
scale the mode-specific travel time distributions by the average travel times. Within the
statistical variation, the resulting distributions appear to be universal. Therefore, the universal
scaled travel time distribution can be considered as a law of human travel behaviour. Apart
from significant deviations for short trips, the curves are well consistent with a canonical
distribution, which can be derived from energy and entropy concepts. The deviations for
short trips are related to the Simonson effect in ergonomics and can be explained by energy considerations
as well (see Sec.~\ref{Sec3}). After relating our energy law for human travel behaviour with previous
approaches in Sec. \ref{Sec4}, we suggest in our outlook, how it could be further tested in the future.

\section{Constant travel time vs. constant energy budgets} \label{Sec1}

In our study, we have investigated the electronically available statistical data of the {\em UK
National Travel Surveys} during the years 1972--1998, which were carried out by the 
{\em Social Survey Division} of the {\em Office of Population Census and Surveys} \cite{eins}. 
The data were gathered on a nation-wide scale. We have used the data records of the so-called
7th day (which does not necessarily mean a weekend day). On this day, short trips were recorded 
as well (in contrast to the other six days). Until 1986, about 25.000 travellers have
been evaluated, later on about 8.000 persons. For this reason, we are facing a worse statistics
than in experimental physics, where the sampling sizes are normally much larger. Nevertheless,
certain features of the data become visible.
\par
\begin{figure}[hptb]
\begin{center}
\hspace*{-2mm}\includegraphics[width=6.5cm, angle=-90]{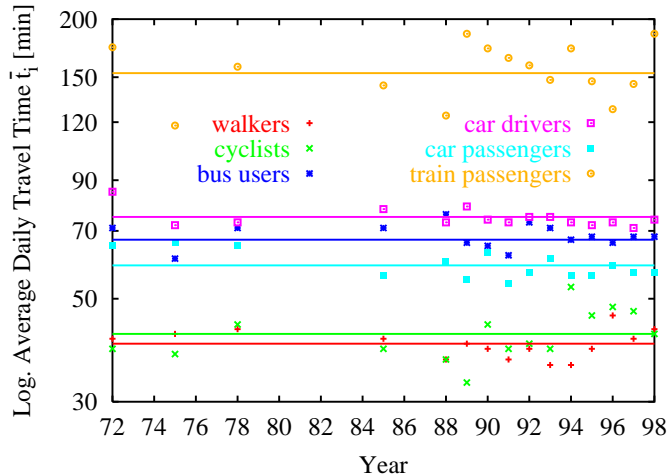}%
\end{center}
\caption{Average travel times for different daily modes of transport in the course of
the years (symbols). Within the statistical variation, the data are mainly compatible with
constant fit curves (solid lines).\label{Fig1}}
\end{figure}

We have investigated the overall daily journey times.
Short and long walks were grouped
together, and the use of London and other stage busses was summed up. Moreover, we 
have focussed on the days
on which, apart from walking, only one mode of transport $i$ has been used. (This is the
case in 80\% of all daily travel.)
When we distinguished the different daily modes of transport, we found that the average
modal travel time $\overline{t}_{i}$ per day and person remained almost
constant over the 27 years of observation
(see Fig.~\ref{Fig1}). More specifically,
the average travel times $\overline{t}_{i}$ were $40
$~min during a day with walking ($i=\mbox{w}$), without any usage
of other means of transport, $42%
$~min for cyclists ($i=\mbox{c}$), $67%
$~min for stage bus users ($i=\mbox{b}$), $75%
$~min for car drivers ($i=\mbox{d}$), $59%
$~min for car passengers ($i=\mbox{p}$), and $153%
$~min for train passengers ($i=\mbox{t}$). The 95\%-confidence intervals
$C_i$ for the slopes $b_i$ of the regression curves were 
$C_{\rm w} = [-0.26;0.16]$,
$C_{\rm c} = [-0.09;0.63]$,
$C_{\rm b} = [-0.45;-0.02]$,
$C_{\rm d} = [-0.576;-0.17]$,
$C_{\rm p} = [-0.32;0.28]$, and
$C_{\rm t} = [-1.27;1.96]$,
i.e. the hypothesis of constant average travel times
corresponding to slopes $b_i = 0$ was predominantly
compatible with the data for most modes over more than 25 years, 
i.e. typical transport and urban planning horizons. From the analysis below, it will become evident why the 
car-related values seem to decrease slightly. We should, therefore, underline that the
analysis in Sec.~\ref{Sec2} can be also applied in situations where the average travel
times (apart from statistical variations) change over the years in a systematic way. 
\par
Note that the average travel times vary by a factor 3.8 between different modes of transport.
This questions the commonly believed hypothesis of a constant average travel time
budget of about 75 minutes a day \cite{fuenf}. Instead, each mode of transport
has its own constant average travel time budget, which calls for a new unifying law of
human travel behaviour. Such a law exists, and it is the
average energy consumed by travel activity which is more or less a mode-independent constant.
We are talking about the energy consumption of the human body, here, not about the
energy consumption by the car, bus, or train.
\par
\begin{table}
\begin{center}
\begin{tabular}{|l|c|c|}
\hline
Activity & Speed & Energy Consumption \\
 & (km/h) & (kJ/min) \\
\hline
\hline
Sitting on a chair &  & 1.5 \\
Standing, relaxed & & 2.6 \\
Standing, restless & & 6.7 \\
\hline
Walking on even path & 4 & 14.1 \\
                                       & 5 & 18.0 \\
\hline
Cycling on even path & 12 & 14.7 \\
\hline
Car, roads & & 4.2 \\
Car, test drive & & 8.0 (5.9--12.6) \\
Car, in city, rush hour & & 13.4 \\
\hline
\end{tabular}
\end{center}
\caption[]{Measured average values of energy consumption per unit time for different kinds of
activities (after \cite{vier}).\label{Tab1}}
\end{table}
To support this idea, let us study values of the energy consumption 
$p_i$ per unit time for certain activities obtained by ergonomic measurements of 
the related O$_{2}$-consumption \cite{vier} (see Table~\ref{Tab1}). The data suggest that the average
daily travel times $\overline{t}_i$ are inversely proportional to the energy consumption rates
$p_i$, i.e.
\begin{equation}
  p_i \approx \frac{\,\overline{E}\,}{\,\overline{t}_i\,} \, .
\label{proportional}
\end{equation}
With an estimated average daily travel energy budget of $\overline{E}\approx 615$~kJ, the  
energy consumption per unit time should amount to $p_{\rm w} = 15.4$~kJ/min for
walking, $p_{\rm c} = 14.6$~kJ/min for cycling, $p_{\rm b} = 9.2$~kJ/min for
bus users, $p_{\rm d} = 8.2$~kJ/min for car drivers, $p_{\rm p} = 10.4$~kJ/min 
for car passengers, and $p_{\rm t} = 4.0$~kJ/min for train passengers.
\par
Comparing this with Table \ref{Tab1}, the energy consumption of 15.4 kJ/min for walking corresponds
to a realistic speed of 1.2 m/s (on an even path and without carrying weight), the 14.6 kJ/min are in harmony
with a cycling speed of 12 km/h, the 8.2 kJ/min for car drivers fit the 8.0 kJ/min of test drive data. The 
elasticity of the energy consumption during driving at roads (with 4.2 kJ/min) and driving in rush hour 
(with 13.4 kJ/min) can explain the apparent slight decrease in average travel time. 
As traffic conditions have obviously become heavier during the last decade, 
driving under these conditions is also 
more stressful, and with that increases the energy expenditure. Hence, under the 
hypothesis of a constant energy expenditure, the travel times are expected to decrease. Similar interpretations
can account for small changes in average travel times for other means of transport.
\par
Finally, the energy consumption of bus, train and car passengers can be 
understood as weighted averages of the energy consumption by walking 
to and fro the bus stop, train station or 
meeting point (15.4 kJ/min) and the energy consumption 
between restless standing with 6.7 kJ/min and sitting on a chair with 1.5 kJ/min. 
Here, one should keep in mind that the actual values for standing and sitting
in a moving means of transport are probably higher, as the body has to perform 
an additional balancing activity.

\section{Derivation of a universal travel time distribution} \label{Sec2}

\begin{figure}[hptb]
\begin{center}
\hspace*{1mm}\includegraphics[width=6.82cm, angle=-90]{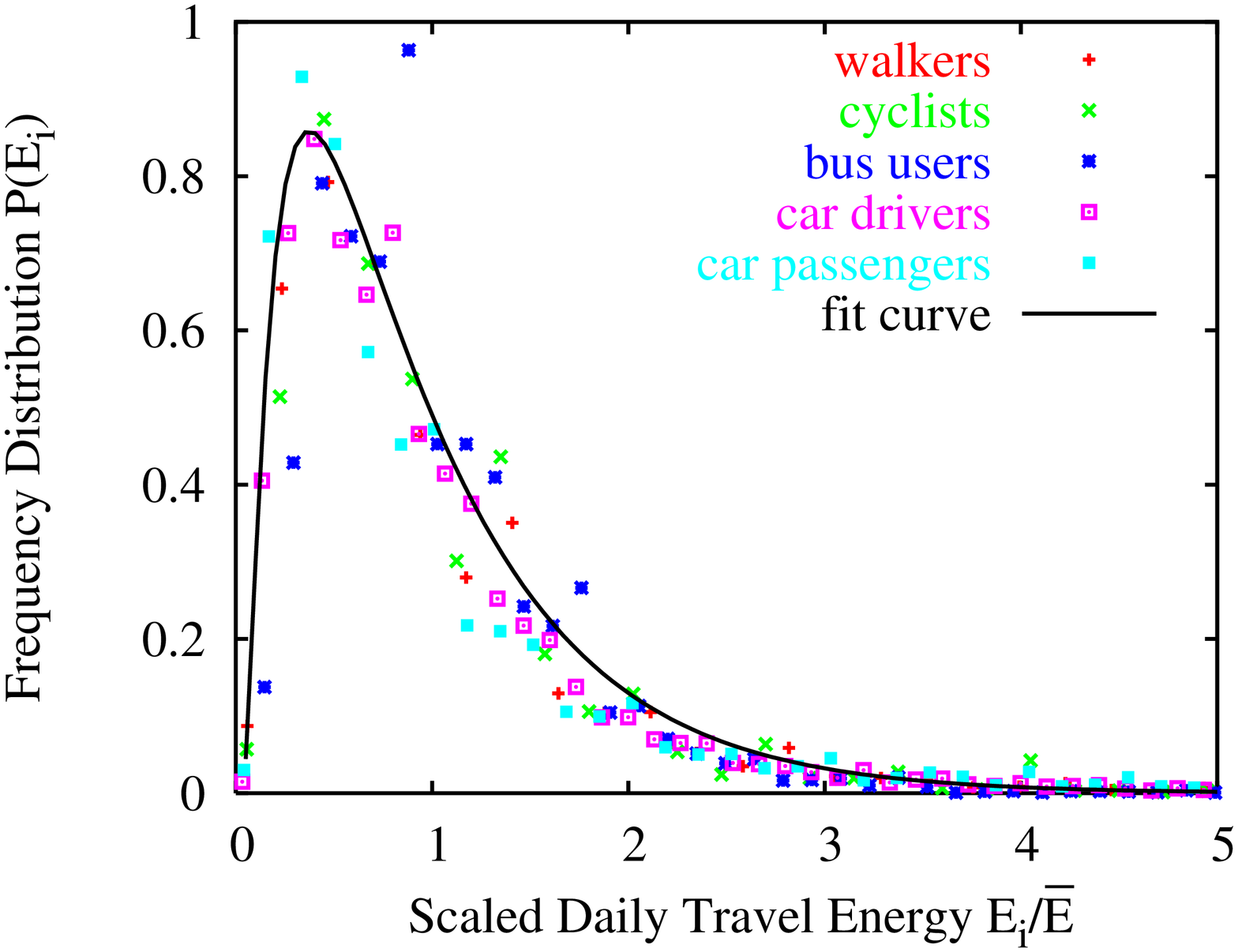}\\
\hspace*{0mm}\includegraphics[width=6.6cm, angle=-90]{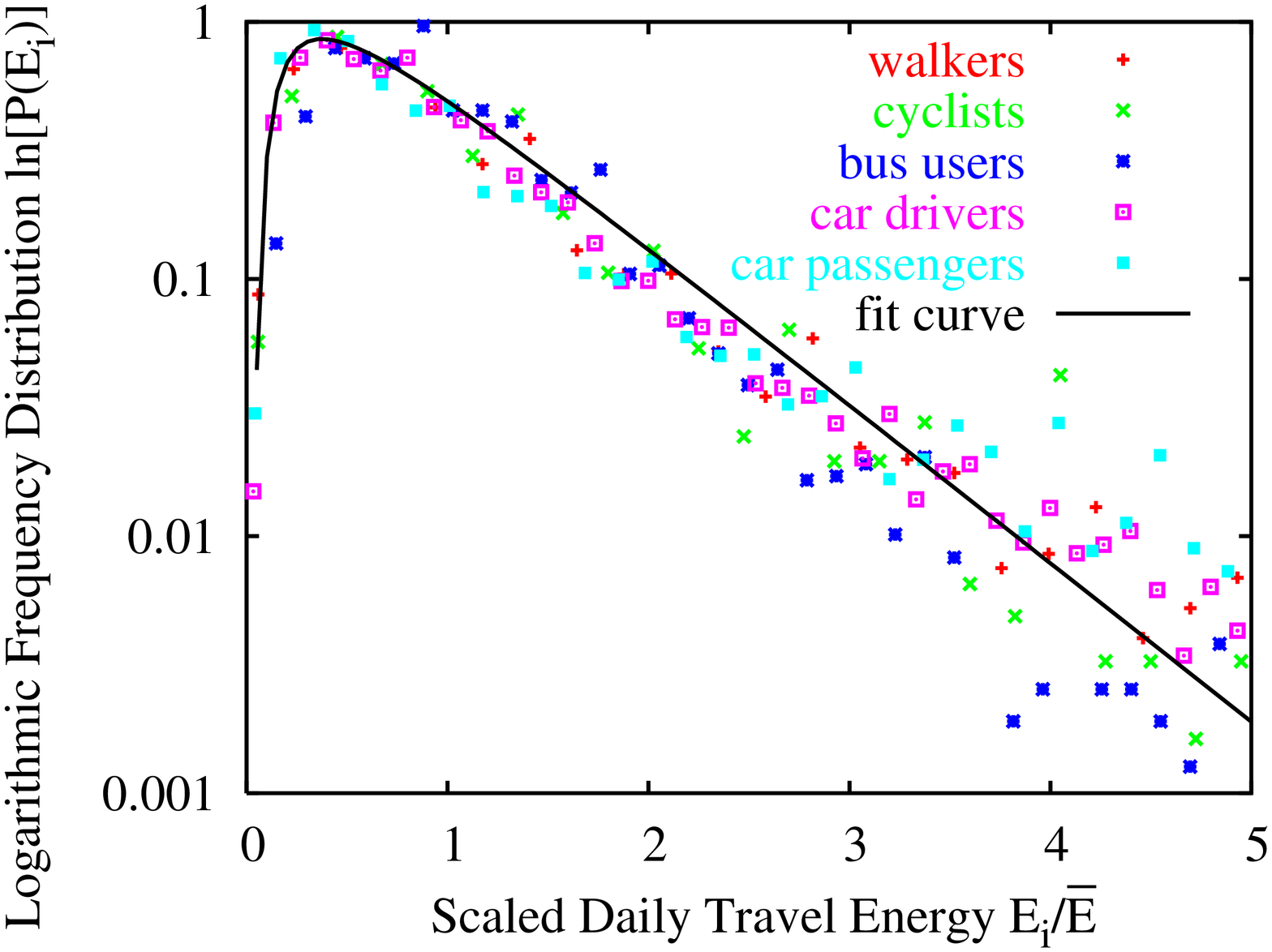}
\end{center}
\caption{Scaled time-averaged travel-time distributions for
different modes of transport in linear (top) and semilogarithmic representation
(bottom). Within the statistical variation (and rounding errors for small frequencies,
which are magnified in the semilogarithmic plot),
the mode-specific data could all be fitted by a universal curve, the travel-energy
distribution (\ref{first}) with $\alpha= 0.2$, $\beta= 0.7$, and normalization constant
$N(\alpha,\beta) = 2.5$. The rail data were not significant because of a large scattering
due to the low number of data.\label{Fig2}}%
\end{figure}

When we scaled the $t$-axis of the modal travel-time distributions
$P_{i}(t)\,dt$ by the average travel times $\overline{t}_{i}$, we discovered
that, within the statistical variation, the resulting distributions collapsed onto
one single curve
\begin{equation}
P'(\tau_{i})\,d\tau_{i}\approx N'\exp(-\alpha/\tau_{i}-\tau_{i}/\beta)\,d\tau_{i} 
\label{one}
\end{equation}
with $\tau_{i}=t/\overline{t}_{i}$, only two fit parameters $\alpha$ and $\beta$, and
the normalisation constant $N=N(\alpha,\beta)$ (see Fig. \ref{Fig2}). This implies a
universal law \cite{zwei} of human travel behaviour which quantitatively reflects 
both, variations of individual travel times and variations between individuals. 
Defining $E_i = p_i t = \overline{E} \tau_i$, this corresponds to a universal travel energy distribution
\begin{equation}
 P(E_i)\,dE_{i}\approx N\exp[-\alpha\overline{E}/E_{i}-E_{i}/(\beta\overline{E})]\,dE_{i} 
\label{first}
\end{equation}
with $N=N'/\overline{E}$. If $\alpha$ was zero, this would correspond to the canonical distribution,
and $\beta$ were 1 in order to guarantee not only the normalization 
condition
\begin{equation}
 \int\limits_0^\infty P(E_i)\,dE_{i} = 1 \, ,
\label{norm}
\end{equation}
but also meet the definition 
\begin{equation}
 \int\limits_0^\infty E_i P(E_i)\,dE_{i} = \overline{E} 
\label{average}
\end{equation}
of the average energy. Because of condition (\ref{average}), $\beta$ has values different from
one, when $\alpha$ is not zero. However, in the semi-logarithmic representation 
\begin{equation}
\ln P(E_{i}) = \ln N-\alpha \overline{E} /E_{i} - E_{i}/(\beta\overline{E}) \, , 
\end{equation}
the term $-\alpha \overline{E}/E_{i}$ is relevant only up to $E_{i} /\overline{E} \approx 0.5$,
while the linear relationship $\ln P(E_{i})=\ln N-E_{i}/(\beta\overline{E})$ dominates
clearly over a wide range of the travel energies $E_{i}$ (see Fig.~\ref{Fig2}). Although the
different modes of transport could, of course, be fitted by separate parameter values, the hypothesis of
identical values is supported by statistical tests. For example, the 95\%-confidence intervals
for the values of $\beta$ are
$\beta_{\rm w} = [0.30;0.74]$,
$\beta_{\rm b} = [0.34;0.95]$,
$\beta_{\rm c} = [0.38;0.71]$,
$\beta_{\rm p} = [0.32;0.71]$, and
$\beta_{\rm b} = [0.56;1.10]$, when we fit $E_i/\overline{E}$ over $\ln P(E_i)$ in the range
$1 \le E_i/\overline{E} \le 6$. 
\par
We suggest the following interpretation: The dominating term 
\begin{equation}
P(E_{i})\,dE_{i}\propto\exp[-E_{i}/(\beta\overline{E})]\,dE_{i}
\end{equation}
corresponds to the canonical energy distribution,
while the term $\exp(-\alpha\overline{E}/E_{i})$ reflects the Simonson effect \cite{drei}. 
It describes the suppression of short trips, a surprising effect, which will have to be explained
in Sec. \ref{Sec3}. 
\par
Based on entropy maximisation, the canonical distribution may be interpreted as 
the most likely distribution, given that the average energy
consumption $\overline{E}$ per day by an ensemble of travellers
is fixed for the area of investigation \cite{Haken}. For the sake of completeness,
we shortly repeat the derivation of this well-known result for a general readership. 
We have to maximise the entropy function \cite{Haken}
\begin{equation}
  - \int dE_i \; P(E_i) \ln P(E_i)
\end{equation}
under the constraints (\ref{norm}) and (\ref{average}).
This is equivalent to maximising
\begin{eqnarray}
 - \int dE_i \; P(E_i) \ln P(E_i) &+& \lambda \bigg[1 - \int dE_i \; P(E_i)\bigg] \nonumber \\
&+& \mu \bigg[\overline{E} -  \int dE_i \; E_i P(E_i) \bigg]
\label{Lagrange}
\end{eqnarray}
and determining the Lagrange multiplyers $\lambda$, $\mu$ so that 
(\ref{norm}) and (\ref{average}) are fulfilled. Functional differentiation of (\ref{Lagrange}) 
leads to the condition
\begin{equation}
 - \ln P(E_i) - 1 - \lambda - \mu E_i = 0 \quad \mbox{or} 
\quad P(E_i) =  \mbox{e}^{ - 1 - \lambda - \mu E_i} \, .
\end{equation}
The parameters $\lambda$ and $\mu$ are now specified in such a way that
the constraints (\ref{norm}) and (\ref{average}) are satisfied. The result for $P(E_i)$ 
corresponds to the canoncial distribution. It agrees well with the investigated data if only
$E_i/\overline{E} = t/\overline{t_i} > 0.5$.

\section{Explanation of the Simonson effect for short trips} \label{Sec3}

Let us now explain the prefactor $\exp(-\alpha\overline{E}/E_{i})$. It reflects that short trips
are less likely to be undertaken, because it is not worth to spend an additional amount of energy
of the order $\alpha \overline{E}$ for the preparation of the trip. 
The quotient between
the average energy spent for the preparation of the trip and the trip itself is $\alpha \approx 0.2$. This value agrees,
by the way, well with the threshold value of accepted detours in route choice behaviour \cite{trails}.
\par\begin{figure}[hptb]
\begin{center}
\hspace*{1mm}\includegraphics[width=6.82cm, angle=-90]{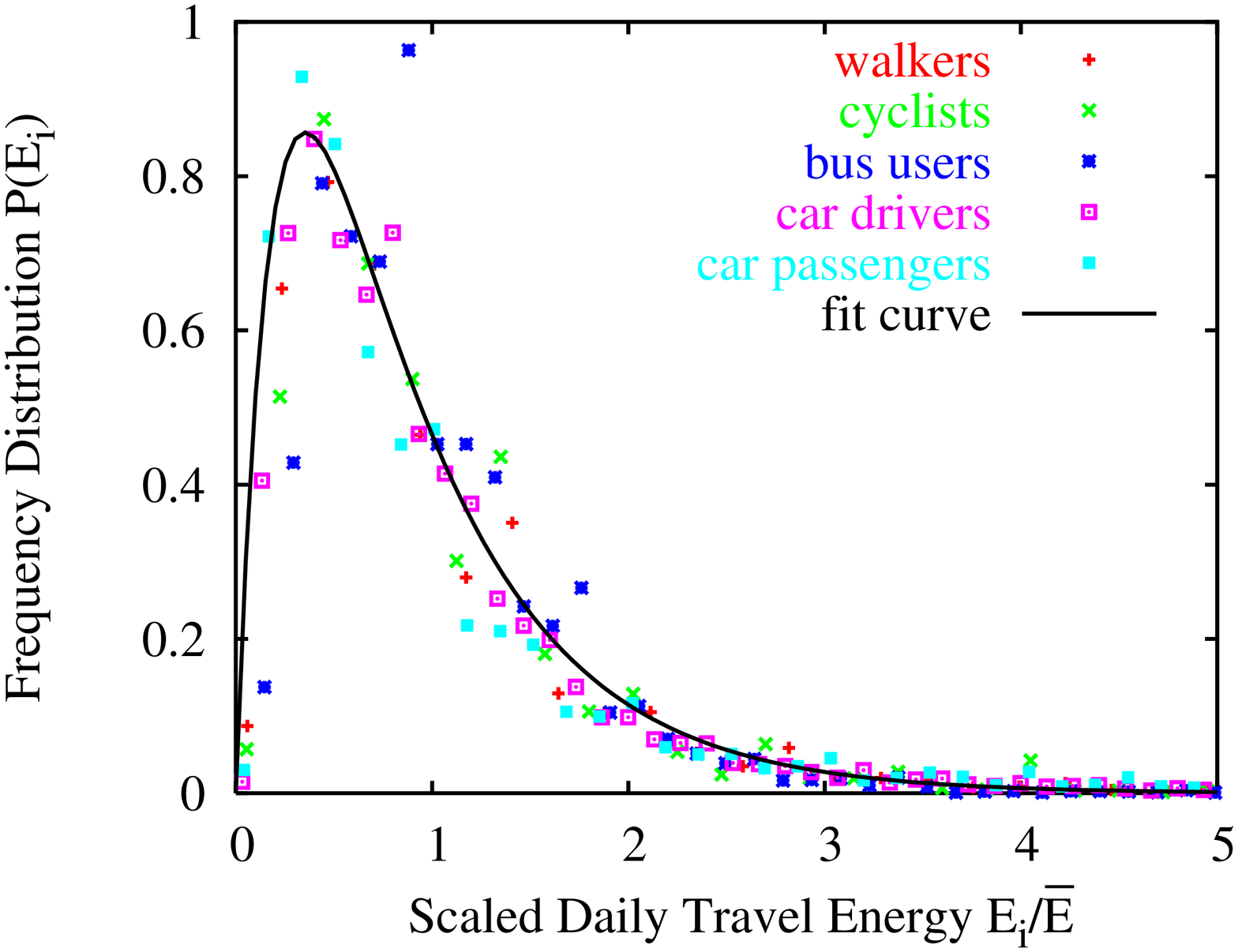}\\
\hspace*{0mm}\includegraphics[width=6.6cm, angle=-90]{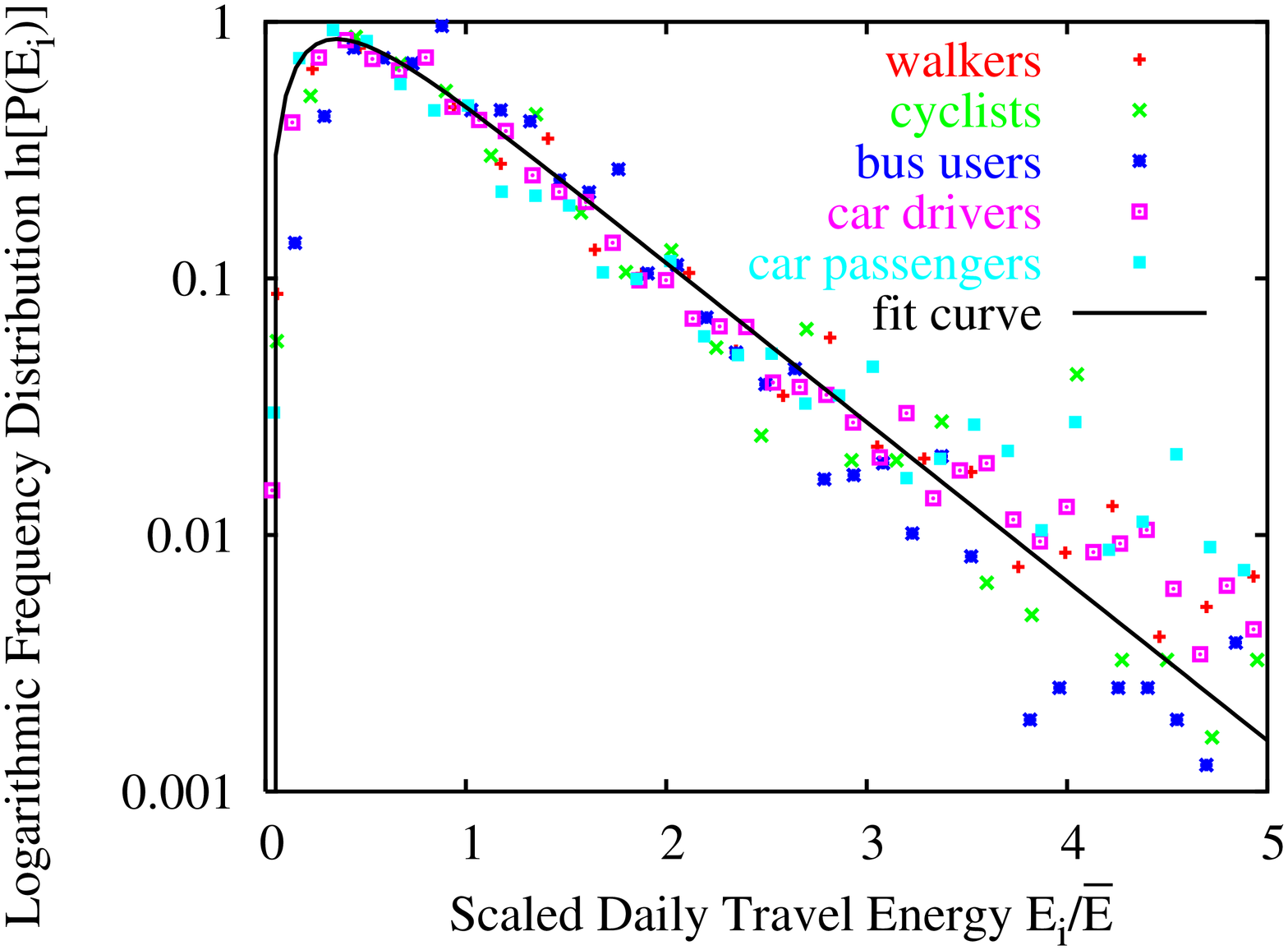}
\end{center}
\caption{Scaled time-averaged travel-energy distributions for
different modes of transport in linear (top) and semilogarithmic representation
(bottom). Within the statistical variation, 
the mode-specific data could all be fitted by the travel-time
distribution (\ref{second}) with the same value of $\beta$ as in Fig.~\ref{Fig2} (i.e. $\beta = 0.7$), 
$\gamma = 3.5$, and normalization constant $N_*(\gamma,\beta) = 2$.\label{Fig3}}%
\end{figure}
We should note that the amount of data does not allow us to decide between this fit function and very similar
ones such as 
\begin{equation}
 P(E_i)\,dE_{i}\approx N_* [ 1 - \exp(-\gamma E_i/\overline{E})] \exp[- E_{i}/(\beta\overline{E})]\,dE_{i} \, ,
\label{second}
\end{equation}
where $\beta$ and $\gamma$ are two fit parameters and $N_*(\gamma,\beta)$ a normalization
constant (see Fig.~\ref{Fig3}). This function
is not so suitable for semi-logarithmic fitting, but it can be theoretically derived. 
For this, let us assume that
the preparation of a trip requires an additional amount of energy $E_0$. 
For similar reasons as outlined in Sec. \ref{Sec2}, the most likely distribution of preparation energy is
\begin{equation}
 P^{\prime\prime} (E_0) \; dE_0 = N^{\prime\prime} \exp[-E_0/\overline{E_0}] \; dE_0  \, ,
\end{equation}
where $\overline{E_0}$ denotes the average energy required for trip preparation.
The trip will not be started, if this energy exceeds a 
certain proportion $\epsilon$ of the energy $E_i$ required for the trip itself. The probability
to start a trip requiring energy $E_i$ is, then, given by
\begin{equation}
  \int\limits_0^{\epsilon E_i} \!\! 
N^{\prime\prime} \exp[-E_0/\overline{E_0}] \; dE_0  
= 1 - \exp(-\epsilon E_i/\overline{E_0}) \, . 
\end{equation} 
Comparing this with Eq. (\ref{second}) establishes the relation $\epsilon/\overline{E_0} = \gamma / \overline{E}$.
\par
We underline that the interpretation of the approach (\ref{second}) basically agrees with the one
behind the simpler formula (\ref{first}). 

\section{Comparison with previous trip distribution models} \label{Sec4}

In this study, we have shown that empirical travel time data suggest the concept of
a constant daily travel energy budget rather than a travel time budget. Not only are the
mode-dependent average travel times inversely proportional to the energy consumption
related to the respective physical activities. Scaling to energy variables also leads to a
universal travel energy distribution, which can be theoretically supported. The distribution
basically corresponds to a canonical one, which can be derived from entropy principles. However,
we found less trips with small travel energies than expected according to the canonical distribution.
This could be explained as effect of the additional energy required to prepare a trip. If the latter exceeds
a certain percentage of the energy required for the trip itself, the trip is unlikely to be
undertaken. In such cases, another means of transportation is possibly selected, or the trip may 
be combined with another trip.
\par
Note that the canonical distribution 
is analogous to the multinomial logit model $P_{i}^{\ast}\propto
\exp(U_{i})$ developed to describe the trip distribution as a function of some utility
function $U_i$ \cite{sechs,MNL,sociodyn}. Therefore, the above theory cannot only give an alternative, physical 
explanation for the applicability of the multinomial logit model. It also allows to extend and
improve previous trip distribution models, which have focussed on variables such as distances and
travel costs \cite{sechs,MNL}. For this purpose, the mode-dependent energy consumption
$E_i$, which our study showed to be an essential variable of human travel behaviour, is to be
taken into account as a (negative) cost term entering the utility function $U_i$. 
\par
Besides, the travel energy concept allows 
one to understand why it is difficult to establish an explanatory principle for all modes 
of transport based on {\em distance.} 
As travelled distance in transport is 
mode dependent, it cannot be regarded as an invariant measure. 
Travel costs, on the other hand, can be assumed to be a determinant for modal choice. 
In qualitative terms, this means for the above approach, that a clear distinction 
can be given between trip distribution and {\em modal} choice, where the former 
determines the physical boundary conditions of daily trip making in which the 
latter takes place. As a consequence, both aspects of travel behaviour can be combined 
in a stringent and complementary way.

\section{Summary and outlook}

In our opinion, the significant progress by this study is the identification of a simple and  universal law 
of human daily travel behaviour after many decades of fit models with multiple fit parameters. 
In contrast to utility functions of classical decision models, which are typically 
based on preferences, our model contains only physical variables such as 
travel times and energies, which are well measurable. It was, therefore, possible 
to critically evaluate our travel distribution model, which resulted in a canonical 
travel energy distribution with a correction term for short trips. In contrast, decision models like the 
multinomial logit model or other trip distribution models are usually presupposed
and used to determine unknown utilities, which are normally not measurable in an independent
way. These models could, therefore, not be verified or falsified in a strict sense. 
Furthermore, it was very surprising to find that physical variables determine
human travel behaviour in a fundamental way and that laws from equilibrium 
thermodynamics apply. Once again, this underlines the
importance of physical concepts for the understanding of traffic and transport.
\par
The main advantage and practical relevance
of the behavioural law (\ref{first}) is its expected long-term validity
under changing conditions. It will, therefore, proof to be important for
urban, transport, and production planning: Previous trip distribution models
had to separately determine various parameter values for each mode of transport,
which was related with large errors in the calibration of the model parameters and did not
allow a theoretical interpretation of travel behaviour. Compared to this, 
the discovered scaling law facilitates improved conclusions about 
trip distributions, modal splits, and induced traffic \cite{sieben}
after the more reliable determination of fewer parameters, 
which are (apart from statistical variations) constant or
systematically changing over typical planning horizons. 
The new concept also contributes to  understanding interactions 
between the temporal evolution of settlement 
patterns and transport systems. Furthermore, it helps to assess the
increase in the  acceptability of public transport, when the
comfort of travel is improved,  to predict the usage of new modes of transport,
and to  estimate potential market penetrations of new travel-related products.
\par
The authors are aware of the problems related with the empirical data, which were characterized by
a large scattering and were affected by 
errors in estimating trips and related travel times. For example, travellers tended to
round travel times to full 5 or 10 minute units (so that these intervals were overrepresented in the data). 
However, presently there are no better data available.
In some sense, the current situation in socio-physics is comparable with the early days of classical physics,
where the data situation was also much poorer than today. But empircally justified
principles such as the proposed travel
energy laws will certainly stimulate further research. If only a small fraction of money spent for
experiments in elementary particle physics was invested into measuring human behaviour, one could
certainly make a fast progress. We do now carry out similar studies for travel time data of other countries,
but there are similar problems with the quality of the data. So, basically, one would have to spend considerably
more money for measuring the data.
\par
In order to further test the proposed travel energy laws, the authors suggest 
the following supplementary investigations: 
First, one could study the travel energy distribution of combined trips, i.e. trips using
different means of transport. The total travel energies
\begin{equation}
 E = \sum_i p_i t_i \, ,
\end{equation}
where $t_i$ denotes the travel time spent with mode $i$, should be distributed according
to the law (\ref{first}) as well, if the combined trip does not require additional energy for its preparation
compared to a one-mode trip (or require long additional waiting times). 
Second, we are confident that the concept of a travel energy budget can
be generalized to a human energy concept. 
For example, we expect that hard-working blue collars tend to attribute a smaller
share of energy to travelling than white collars. We also predict that elderly people make shorter trips on average,
because travelling is more exhausting for them. This could be investigated by distinguishing different
groups (subpopulations) of travellers. These questions shall be addressed by future publications.

\subsection*{Acknowledgments} 

R.K. would like to thank Mike McDonald, Hermann Knoflacher, 
Robin Stinchombe, Denis Pollney, David K. Anthony and Mark Brackstone for their 
support and/or inspiring discussions.
D.H. is grateful for comments by Tamas Vicsek,
for the warm hospitality at the Collegium Budapest.

\end{document}